\documentclass[12pt]{article}

\usepackage{dcolumn}
\usepackage{bm}
\usepackage[latin1]{inputenc}
\usepackage[spanish,english]{babel}
\usepackage{amsfonts}
\usepackage{amssymb}
\usepackage{graphicx}
 \usepackage{setspace}
 \usepackage{verbatim} 
 
\usepackage{color}
\usepackage{soul}

\usepackage[margin=3cm]{geometry}

\newcommand{\be}{\begin{equation}}
\newcommand{\ee}{\end{equation}}
\newcommand{\bea}{\begin{eqnarray}}
\newcommand{\eea}{\end{eqnarray}}

\begin{document}

\begin{titlepage}

\begin{center}
{\large \bf Gravity and handedness of photons} 
\end{center}

\begin{center}
Ivan Agullo\footnote{agullo@lsu.edu}\\ 
{\footnotesize \noindent {\it {Department of Physics and Astronomy, Louisiana State University, Baton Rouge, LA 70803-4001, USA}}   }\\

Adrian del Rio\footnote{adrian.rio@uv.es} and  Jose Navarro-Salas\footnote{jnavarro@ific.uv.es}\\

{  \footnotesize \noindent {\it Departamento de Fisica Teorica and IFIC, Centro Mixto Universidad de Valencia-CSIC. Facultad de Fisica,  Burjassot-46100, Valencia, Spain.
} }

\end{center}

\begin{abstract}

Vacuum fluctuations of quantum fields are altered  in presence of a strong gravitational background, with important physical consequences.  We argue that a non-trivial spacetime geometry  can act as an optically active medium for  quantum  electromagnetic radiation,  in such a way that the state of polarization of radiation changes in time,  even in  the absence of  electromagnetic sources.  This is a quantum effect, and is a consequence of an  anomaly related to the classical invariance under electric-magnetic duality rotations in Maxwell theory.\\

\end{abstract}

\vspace{3cm}

\begin{center}{\it First Award in the 2017 Essay Competition of the Gravity Research Foundation.} \end{center}
\end{titlepage}

The presence of a  gravitational background introduces non-trivial  effects in the propagation of electromagnetic radiation, with  fundamental  implications. The gravitational redshift and the deflection of  light rays by massive bodies  played a pivotal role in the birth of general relativity  \cite{Pais}, and are of great importance in many areas of gravitation, cosmology, and astrophysics. 
 It is also well-known that remarkable new features appear in the physics of quantum fields  when they propagate under the influence of gravity. 
Renowned examples are the spontaneous creation of quanta by the expanding universe \cite{parkerthesis}, the thermal radiation by black holes produced by gravitational collapse \cite{hawk1}, and the generation of primordial  density perturbations by inflation \cite{inflation2}. In this essay, we discuss a new   effect on quantum electromagnetic fields propagating in curved spacetimes. 

Our discussion  resembles the study of the chiral anomaly for fermions.
 Recall that the theory of a zero-mass, charged spin-$1/2$ fermion satisfying the Dirac equation  in Minkowski spacetime,  $\gamma^{\mu}\partial_{\mu}\Psi=0$, is invariant under chiral transformations: $\Psi\rightarrow e^{i \theta \, \gamma_5}\Psi$. Noether's theorem then implies  that the so-called axial-vector current, $j^{\mu}_5=\bar \Psi \gamma^{\mu}\gamma_5\Psi$, is  conserved; $\partial_{\mu}j^{\mu}_5=0$. This classical symmetry is present even if we let fermions  interact with a  classical   electromagnetic field $F_{\mu\nu}$. But it is well-known that  quantum fluctuations of the Dirac field lead to an anomaly \cite{ABJ}; the vacuum expectation value of   $\partial_{\mu}j^{\mu}_5$ does not vanish
\be\label{chiralanomaly} \langle \partial_{\mu}j^{\mu}_5\rangle = - \frac{q^2 \hbar}{8 \pi^2} F_{\mu\nu}{^{\star}F}^{\mu\nu} =- \frac{q^2 \hbar}{2 \pi^2}  E_{\mu}B^{\mu}\, , \ee
where $q$ is the charge of the fermion, and $^{\star}F^{\mu\nu}=1/(2\sqrt{-g})\epsilon^{\mu\nu\alpha\beta}F_{\alpha\beta}$ is the dual of $F_{\mu\nu}$. In the last  equality we have written the Lorentz invariant quantity $F_{\mu\nu}{^{\star}F}^{\mu\nu}$ in terms of the electric and magnetic parts of $F_{\mu\nu}$.  This anomaly has important physical implications, e.g.\ it provides the mechanism to explain the pion decay to two photons $\pi^0\rightarrow \gamma\gamma$. 

With this result in mind, we now analyze a different physical situation: the theory of a source-free electromagnetic field propagating on a classical gravitational background. The corresponding Maxwell action is  invariant under duality transformations \cite{deser-teitelboim}: $F_{\mu\nu}\rightarrow F_{\mu\nu}\, \cos{\theta}+ \, ^{\star}F_{\mu\nu}\, \sin{\theta}$. In terms of  the vector potential, the infinitesimal transformation reads $\delta A_{\mu}=\theta \, Z_{\mu}$, where $Z_{\mu}$, when evaluated on-shell, is the potential of the dual field $^{\star}F_{\mu\nu}=\nabla_{\mu}Z_{\nu}-\nabla_{\nu}Z_{\mu}$. Again, Noether's techniques provide  a  conserved current $j^{\mu}_D$, whose expression on-shell is given by \cite{AdrNS16} 
\be \label{jD} j^{\mu}_D = \frac{1}{2}\left(A_{\nu}\, ^{\star}F^{\mu\nu}-F^{\mu\nu}Z_{\nu}  \right)  \ . \ee
 $Q_D$, the spatial integral of the time component of $j^{\mu}_D$ on any  hyper-surface, is a conserved quantity of the classical theory in arbitrary spacetime geometries. In Minkowski spacetime, $Q_D$ represents the difference in  amplitude between  right and left polarized components of the electromagnetic field --- this is  the $V$-Stokes parameter. 
 This conservation law  says  that the state of polarization of light is  a constant of motion.

     The electromagnetic duality is indeed a  remarkable symmetry. 
    It suggested  the existence of magnetic charges as a way to  retain   the symmetry in presence of electromagnetic sources. 
    We argue below, however, that  the gravitational interaction   spoils the symmetry, even without  sources, once quantum fluctuations of the electromagnetic field are taken into account.

That  the transformation law for the basic variables $A_{\mu}$ looks complicated---non-local in fact---indicates that the free Maxwell theory is more naturally formulated in terms of other variables, namely self- and anti-self dual fields.  We will reformulate the theory in terms of these variables, and in doing so we will see an interesting analogy with the dynamics of fermions and the associated chiral symmetry. Powerful techniques developed for fermions will then become available to study the electromagnetic duality in the quantum theory. 
We focus first on Minkowski spacetime, and later allow arbitrary gravitational fields. 

Define the complex fields $\vec H_{\pm}:= \frac{1}{2}[ \vec E\pm i\vec B]$. It is easy to see that they transform under duality by a simple phase, $\vec H_{\pm}\to e^{\mp i \theta} \vec{H}_{\pm}$. $\vec{H}_+$ and $\vec{H}_-$ are called the self- and anti-self-dual parts of the electromagnetic field. They transform under the $(1,0)$ and $(0,1)$ representations of the Lorentz group, respectively.  Define the associated vector potentials $\vec H_{\pm}=: \pm i\, \vec \nabla \times \vec A_{\pm}$.   The Maxwell equations, in  the radiation gauge, read
\be
   \vec \nabla \times \vec A_{\pm}  =\pm  i \, {\partial_t} \vec A_{\pm}\ , \hspace{0.3cm} \vec \nabla\cdot \vec A_{\pm}=0. \label{MaxwellA}
 \ee
These equations can be derived either  by directly writing Maxwell's equations $d F=d\,^{\star}F=0$ in terms of $\vec A_{\pm}$, or as  Hamilton's equations for the Maxwell Hamiltonian 
 using  $2\vec A_{-}$ and $-\vec{H}_+$---or equivalently  $2\vec A_{+}$ and $-\vec{H}_-$---as canonically conjugate variables. 

The four equations (\ref{MaxwellA}) for $\vec A_{+}$ can be written more compactly as 
$(\alpha^{a})^{b}_{\hspace{0.15cm}i} \partial_{a} A^i_+=0 $, where the components of the $(\alpha^{a})^{b}_{\hspace{0.15cm}i}$  matrices are easily extracted from (\ref{MaxwellA})---the spacetime indices $a,b$ run from 0 to 3, and the internal index $i$ runs from  1 to 3. The equations for the anti-self-dual potential $\vec A_{-}$ are   the  complex-conjugates of the equation for $\vec A_{+}$. Both  sets of equations can be combined in
\bea
\beta^{a}\partial_{a}\Psi(x) = 0\ , \label{diracmaxwell} 
\eea
where we have defined 
\bea
\Psi \equiv  \left( {\begin{array}{c}
    A_+^{\ i}  \\
 A_{-\, i}\\
  \end{array} } \right) \ , \hspace{0.6cm}     \beta^{a} \equiv i \left( {\begin{array}{cc}
 0 &  (\bar \alpha^{a})_{b}^{\ i}  \\
 - (\alpha^{a})^{b}_{\ i} & 0 \\
  \end{array} } \right)      \ .
\eea 
 (The bar denotes complex conjugation). The $\beta^{a}$-matrices satisfy the following (anti-) commutation properties: $\bar \beta^{(a} \beta^{b)}  =  - \eta^{ab}\, \mathbb  I \ , \ 
 \bar \beta^{\lbrack a} \beta^{b\rbrack}   =  2$ diag$\left( ^{+}\Sigma^{a b} , ^{-}\Sigma^{ab}  \right)$,
where $\eta_{ab}$ is the Minkowski metric and $^{\pm}\Sigma^{a b}$ are the generators of the $(1,0)$ and $(0,1)$ representations of the Lorentz group, respectively. $\alpha^{a}$ and $\beta^{a}$ are the spin-1 analog of the Pauli $\sigma^{a}=(I,\vec{\sigma})$ and  Dirac $\gamma^{a}$ matrices, respectively. 
Duality transformations are now written as 
\bea
\Psi \rightarrow e^{i\theta \beta_5}\Psi\, = \left( {\begin{array}{c}
 e^{-i\theta} A_+^i \\
e^{i\theta}     A_{-\, i} \\
  \end{array} } \right)\ , \label{3.102}
\eea
where  $ \beta_5 \equiv  \frac{i}{16} \epsilon_{\mu\nu\sigma\rho}
\bar \beta^{\mu}\beta^{\nu} \bar \beta^{\sigma}  \beta^{\rho}  =$ diag$ \left (
  -\mathbb I_{3\times 3} , \mathbb I_{3\times 3} \right)$.  
   
The space of solutions of (\ref{diracmaxwell}) is spanned by transverse monochromatic waves oscillating with positive and negative frequencies. The relation between self-duality and helicity is as follows. Self-dual monochromatic waves have (negative) positive helicity for (negative) positive frequency modes, while the opposite occurs for anti-self dual solutions. This is  analog to the relation between chirality and helicity for  fermions. 

The generalization to curved spacetimes is straightforward. The Minkowski metric $\eta_{ab}$ and the ordinary derivative $\partial_a$ are replaced by the curved metric tensor $g_{\mu\nu}$ and the associated covariant derivative $\nabla_{\mu}$, and the curved spacetime $\alpha$- and the $\beta$-matrices are obtained from the flat spacetime ones  by using an orthonormal tetrad or vierbein, $ (\alpha^{\mu})^{\nu}_{\hspace{0.15cm}i} (x)= e^{\mu}_a(x)\,  e^{\nu}_b(x)\,  (\alpha^{a})^{b}_{\hspace{0.15cm}i}$.

We now explore the duality symmetry in the quantum theory. We will follow the functional-integral strategy used in  \cite{Fujikawa} to establish the chiral anomaly of fermions. 
 The important question is whether the  measure of the path integral respects the  symmetry of the action. The calculation follows steps similar  to  the fermionic case, after replacing  all the structures associated to the $(1/2,0)$ and $(0,1/2)$ representations of the Lorentz group  with  the $(1,0)$ and $(0,1)$ ones, and taking care of additional subtleties arising from the gauge freedom. The reader is referred to \cite{AdrNS16} for details. In spite of the differences, the result is  remarkably similar to the fermionic chiral anomaly. Namely, quantum fluctuations spoil the classical symmetry and  the expectation value of the divergence of (\ref{jD}) ---for any  vacuum state--- becomes
\bea \label{anomalyD}
\left< \nabla_{\mu}j_D^{\mu} \right>= \frac{\hbar}{24 \pi^2}R_{\mu \nu \lambda \sigma} {^{\star}R} ^{\mu \nu \lambda \sigma}=\frac{2\hbar}{3 \pi^2} \, E_{\mu\nu}B^{\mu\nu},  \
\eea 
where in the last equality we have written the Chern-Pontryagin  topological density $R_{\mu \nu \lambda \sigma} {^{\star}R} ^{\mu \nu \lambda \sigma}$ in terms of the electric and magnetic parts of the Weyl tensor. Compare this expression with (\ref{chiralanomaly}). As a consequence, the duality charge $ Q_D  $ is not conserved in general spacetimes, and its  change between any two instants is given by 
\bea
\Delta Q_D= \frac{2\hbar}{3 \pi^2} \int_{t_{1}}^{t_{2}}\int_{\Sigma} d^4 x \sqrt{-g} \, E_{\mu\nu}B^{\mu\nu}  \ . \label{qanomaly}
\eea

An analog of this effect arises for fermions in the creation of  pairs from the vacuum by strong electric fields. In this situation the presence of a magnetic field would induce a non-zero net chirality on the particles created, as predicted by (\ref{chiralanomaly}) \cite{Nielsen-Ninomiya}. Likewise, apart from gravitational tidal forces produced by $E_{\mu\nu}$,  frame dragging effects, described by $B_{\mu\nu}$, are necessary to induce net polarization on the created photons.

To illustrate this, consider  the process of collapse of a neutron star into a Kerr black hole. In the vicinity of the region where the event horizon will form, the gravitational field is strongly changing and {\em spontaneous}  creation of photons will occur. Our result indicates that the  created photons, when measured far from the star, will carry a net polarization given by $\Delta Q_D$. Numerical simulations indicate that for a neutron star of $M=1.73$ solar masses and angular momentum $J=0.36 M^2$   the collapse produces around 30 photons per second more with one circular polarization than the other. This process has no classical counterpart and  is different from the standard, late-time Hawking radiation, which does not contribute to $\Delta Q_D$.\footnote{  In fact, for  an exact Kerr geometry expression  (\ref{qanomaly}) yields  zero. Therefore,  $\Delta Q_D$ comes  from the (transient) process of collapse, in contrast to the Hawking effect, which is associated with the final, stationary black hole configuration.}Although this number is small---given the short duration of the gravitational collapse---it is significant if we compare it with the  $\approx20$ total photons per second,  steadily emitted by the formed black hole via Hawking radiation, with net polarization equal zero \cite{DNPage}. It is expected that  $\Delta Q_D$ becomes significantly larger in more violent processes, as for instance the collision and merger of  two black holes as the ones observed by LIGO \cite{LIGO}.  But more importantly, the existence of spontaneous creation of photons implies that the {\em stimulated} counterpart must exist. Therefore, electromagnetic radiation traveling in spacetimes with a  non-zero  value of (\ref{qanomaly}), such as the ones mentioned above, will experience a change in its net polarization.  This may have implications for CMB photons; their propagation through the large scale structure  would not only bend their trajectories but also affect their polarization. \\

{\it Acknowledgments}. We are very grateful to N. Sanchis-Gual and J. A. Font for numerical assistance in the black hole collapse, and to K. Runnels for a careful reading of the manuscript. This work was  supported  by the  Grants No.\ FIS2014-57387-C3-1-P, and No.\ MPNS of COST Action No. MP1210,  the Severo Ochoa program SEV-2014-0398 and NSF Grants No. PHY-1403943 and No. PHY-1552603. 
A. d. R. is supported by the Spanish FPU Ph.D. fellowship.

\end{document}